\documentclass[journal]{IEEEtran}

\ifCLASSINFOpdf
\else
   \usepackage[dvips]{graphicx}
\fi
\usepackage{url}

\usepackage[fontsize = 9.6pt]{scrextend}
\usepackage{xcolor}
\usepackage{changes}
\usepackage{graphicx}
\usepackage{multirow}
\usepackage{amsmath}
\usepackage{float}
\usepackage{color}
\usepackage{amsfonts}
\usepackage{balance}

\definechangesauthor[name={fp},color=blue]{fp}
\usepackage{hyperref}
\begin{document}

\title{Key Frame Mechanism For Efficient Conformer Based End-to-end Speech Recognition}
\author{Peng Fan, Changhao Shan, Sining Sun, Qing Yang, Jianwei Zhang

\thanks{\textit{Corresponding author: Sining Sun.}}
\thanks{\textit{Peng Fan and Changhao Shan contributed equally to this work.}}
\thanks{This work was supported by Key R\&D Projects in Sichuan Province under Grants 2022YFG0261.}
\thanks{Peng Fan is with the National Key Laboratory of Fundamental Science on Synthetic Vision, Sichuan University, Chengdu 610065, China(email: fanpeng2023@hotmail.com)}
\thanks{Changhao Shan, Sining Sun and Qing Yang are with Du Xiaoman Financial, Beijing 100089, China (email: changhaoshan@gmail.com,  ssning2013@gmail.com)}
\thanks{Jianwei Zhang is with the College of Computer Science, Sichuan University, Chengdu 610065, China (email: zhangjianwei@scu.edu.cn)}
}

\markboth{Journal of \LaTeX\ Class Files, Vol. 14, No. 8, August 2015}
{Shell \MakeLowercase{\textit{et al.}}: Bare Demo of IEEEtran.cls for IEEE Journals}
\maketitle

\begin{abstract}
Recently, Conformer as a backbone network for end-to-end automatic speech recognition achieved state-of-the-art performance. The Conformer block leverages a self-attention
mechanism to capture global information, along with a convolutional neural network to capture local information, resulting in improved performance. However, the Conformer-based model encounters an issue with the self-attention mechanism, as computational
complexity grows quadratically with the length of the input sequence. Inspired by previous Connectionist Temporal Classification (CTC) guided blank skipping during decoding, we introduce intermediate CTC outputs as guidance into the downsampling procedure of the Conformer encoder. We define the frame with non-blank output as key frame. Specifically, we introduce the key frame-based self-attention (KFSA) mechanism,
a novel method to reduce the computation of the self-attention mechanism using key frames. The structure of our proposed approach comprises two encoders. Following the initial encoder, we introduce an intermediate CTC loss function to compute the label frame, enabling us to extract the key frames and blank frames for KFSA. Furthermore, we introduce the key frame-based downsampling (KFDS) mechanism to operate on high-dimensional acoustic features directly and drop the frames corresponding to blank labels, which results in new acoustic feature sequences as input to the second encoder. By using the proposed method, which achieves comparable or higher performance than vanilla Conformer and other similar work such as Efficient Conformer. Meantime, our proposed method can discard more than 60\% useless frames during model training and
inference, which will accelerate the inference speed significantly. This work code is available in \href{https://github.com/scufan1990/Key-Frame-Mechanism-For-Efficient-Conformer}{(https://github.com/scufan1990/Key-Frame-Mechanism-For-Efficient-Conformer)}
\end{abstract}


\IEEEpeerreviewmaketitle

\section{Introduction}

\IEEEPARstart{R}{ecently}, the transformer model incorporating self-attention mechanism has emerged as a leading approach in natural language processing (NLP), demonstrating state-of-the-art results across various sequence-to-sequence tasks \cite{vaswani2017attention}. The Conformer \cite{gulati2020conformer}, a variant of the transformer, which incorporates a self-attention mechanism and convolutional neural network (CNN), has emerged as a popular backbone network for automatic speech recognition (ASR). 

According to the decoder's type, end-to-end ASR model can be divided into Connectionist Temporal Classification (CTC) based
model, attention-based Encoder-Decoder (AED) model, and  RNN-Transducer (RNN-T) model \cite{2016Listen,2018Speech,graves2006connectionist,graves2012sequence,fan2021speech,zhang2021tiny,graves2014towards,he2019streaming,battenberg2017exploring}.
{In order to align the input and output sequences, a blank symbol is introduced in CTC and RNN-T models.  However, CTC or RNN-T model shows peaky posterior property, and ignoring blank frames' posterior during decoding will not
introduce additional search errors~\cite{chen2016phone,zhang2021tiny}.  This interesting finding attracts more and more researchers to further accelerate CTC or RNN-T model's inference speed by ignoring blank symbols with different strategies. For example, Tian et al.~\cite{tian2021fsr} proposed Fast-Skip Regularization (FSR) method, which co-trains CTC and RNN-T. During transducer decoding, blanks predicted by the CTC model will be ignored. In~\cite{xu2022multiblank}, Xu et al. proposed multi-blank transducer, which introduced big blank symbols. Big blank symbols are blanks with explicitly defined durations. More than one blank frames can be ignored when the big blank is predicted. In~\cite{yang2023blank}, the authors focused on encouraging more blanks during CTC inference so that more frames can be skipped. 
\textcolor{black}{Jonathan et al. introduced amortized neural networks in the RNN-T-based ASR model to reduce computational costs and latency}~\cite{macoskey2021amortized}.

}

\label{sec:guidelines}
\begin{figure*}[ht!]
 	\centerline{\includegraphics[width=0.95\linewidth]{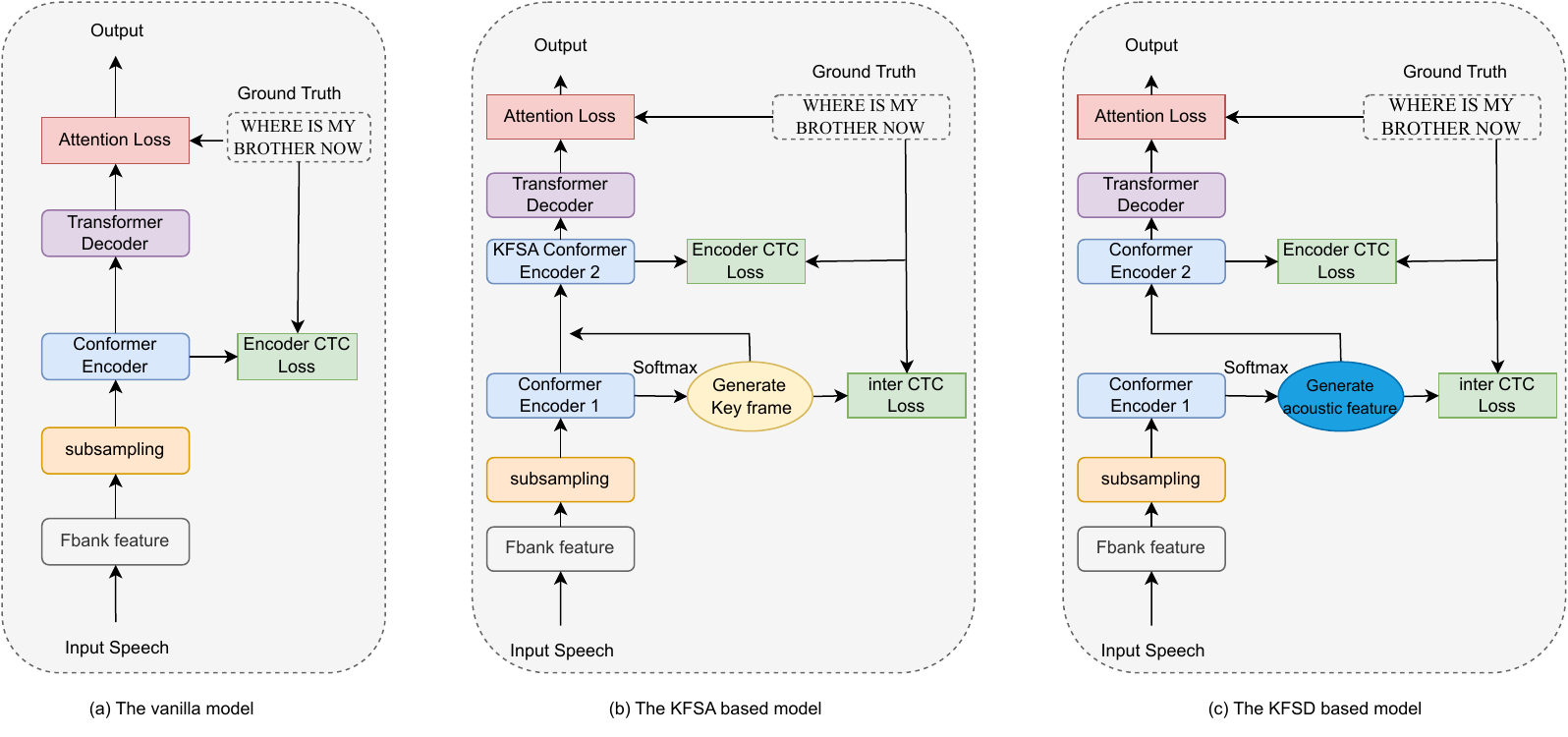}}
 	\caption{The overall architecture of the vanilla Conformer-based AED model (a), the proposed KFSA-based model (b), and the proposed KFDS-based model (c). }\label{fig1}
\end{figure*}

The above-mentioned work can reduce decoding time by skipping blank outputs, but each frame has to be processed by the encoder. Conformer-based encoder which relies on self-attention mechanisms is challenged by the issue of quadratic growth
in computational complexity with the increase of input length. Therefore, further acceleration could be obtained if more frames can skip over the encoder in advance. To this end, some work tried to reduce self-attention calculation, such as Prob-Sparse attention mechanism~\cite{wang2023accelerating} and HybridFormer~\cite{10096467}. Other work adapted downsampling strategies along the time axis such as Efficient Conformer~\cite{burchi2021efficient} and SqueezeFormer~\cite{kim2022squeezeformer}.  Both Efficient Conformer and SqueezeFormer used a convolution downsampling module with a uniform subsampling strategy in the middle of the Conformer encoder. Subsampling feature embedding vectors uniformly along temporal dimensions can reduce unnecessary computational overhead and information redundancy. However, subsampling uniformly may lead to missing crucial information because the distribution of speech signals is pretty complex. 

In this paper, we mainly focus on how to select crucial frames efficiently and effectively for the Conformer encoder. Inspired by previous CTC-guided blank skipping  during decoding, we introduce CTC outputs into the downsampling procedure of the Conformer encoder. The most related work with our paper is~\cite{wang2023accelerating}, which proposed to do frame reduction in the middle of the RNN-T encoder using co-trained CTC guidance. In~\cite{wang2023accelerating}, the authors assumed that if an encoder embedding frame is classified as a blank frame by the CTC model, it is likely that this frame will be aligned to blank in the RNN-T model. Different from~\cite{wang2023accelerating}, in this work, an intermediate CTC loss~\cite{interctc} is attached to an intermediate layer in the Conformer encoder. Frame reduction will be guided by intermediate CTC outputs. The frames with a non-blank output of intermediate CTC should contain much more semantic information than the blank ones, which should contribute much more to the final performance. Therefore, we define the frame with non-blank output as \textit{key frame}. 

Specifically, in this paper, we propose a key frame-based self-attention (KFSA) mechanism and a key frame-based downsampling (KFDS) mechanism. According to the location of the intermediate CTC, the whole Conformer encoder can be divided into two parts. The first part consists of vanilla Conformer layers. KFSA or KFDS mechanism is applied in the second part of the encoder, where Fig. 1 
gives more details. Key frames are selected from the output of the first encoder based on the intermediate CTC prediction. For KFSA mechanism, only key frames are considered as they contain global semantic information. Furthermore, we also explore the effect of local context by attending to adjacent frames of the current key frame during attention calculation. In practice, the KFSA mechanism can be implemented by using a well-designed mask. KFSA mechanism reduces the complexity of attention calculation from $O(dT^2)$ to $O(dU^2)$ in theory, where $T$ is the frame number of acoustic feature sequence, $U$ is the length of label sequence, and $d$ is the attention dimension. In general, for speech recognition tasks, $T\gg  U$. Furthermore, we also utilize intermediate CTC output to guide the downsampling of the second Conformer encoder, which is the proposed KFDS mechanism. Our proposed KFDS method can reduce at least 60\% frames and even better ASR performance compared with vanilla Conformer.

\section{METHODLOOGY}



In this section, we will give more details on our proposed KFSA and KFDS methods.

\begin{figure}[h!]
	\centerline{\includegraphics[width=0.5\linewidth]{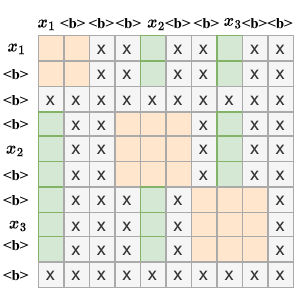}}
	\caption{The KFSA mask. Grey areas are masked out. The light orange area can be used to attend to other light orange areas (local information) and light green areas (global information)}\label{fig_mask}
\end{figure}

\subsection{The self-attention mechanism}
The self-attention mechanism is a key part of Transformer, which has a strong global feature learning ability \cite{vaswani2017attention}.
Given input feature matrix $X \in \mathbb{R}^{T \times d_{x}}$, where $T$ and $d_x$ are frame number of acoustic feature sequence and the dimension of input respectively. The self-attention mechanism first projects $X$ into query matrix $Q = XW_q$, key matrix $K = XW_k$ and value matrix $V = XW_v$, where $\{Q, K, V\} \in \mathbb{R}^{L \times d}$ and $W_q$, $W_k$, $W_v$ are projection matrices with appropriate shapes, where $d$ is the Query and Key's dimension. We compute the dot products of the query with all keys, divide each by $\sqrt{d}$, and apply a softmax function to obtain the weights on the values. Then self-attention mechanism can be represented by

\begin{equation}
   \operatorname{Attention}(Q, K, V)=\operatorname{softmax}\left(\frac{Q K^\mathsf{T}}{\sqrt{d}}\right) V.
\end{equation}

\subsection{The KFSA mechanism}

The proposed KFSA mechanism aims to reduce frames that need to attend the attention calculation with extra guidance. Only key frames that contribute more to the ASR performance will be selected. Fig 1. (b)  depicts the architecture of the proposed KFSA mechanism.  In Fig. 1 (b), Conformer encoder 1 is joint training by intermediate CTC loss. The output of intermediate CTC is used to generate mask $M\in \{0,1\}^{T\times T}$ for attention calculation. Let $C\in \mathbb{R}^{T\times V}$ be the output of intermediate CTC and $V$ is the vocabulary size. Let $\mathbb{P}$ be the non-blank position set defined by Eq.~\ref{eq_p}. Note that continuous duplicate non-blank outputs are removed from this set.  
\begin{equation}\label{eq_p}
    \mathbb{P}=\{t|\max{C_t} \not=blank\}
\end{equation}

Suppose that  $t_p$ is one of elements in $\mathbb{P}$, that is $t_p\in \mathbb{P}$. For any $t_1$ and $t_2$, the corresponding mask can be obtained by Eq~\ref{eq_mask}, where $w$ is the local context width. If $w=0$, no local context is used. 
\begin{equation}\label{eq_mask}
    M_{t_1, t_2}=
    \begin{cases}
        1, &if\ \left|t_1-t_p\right|\le w\ or\ t_2 \in \mathbb{P} \\
        0, &else
    \end{cases}
\end{equation}

Fig.~\ref{fig_mask} gives an example mask for KFSA mechanism with $w=1$. For blank frames, which do not locate in any local context window of key frame, will be skipped from self-attention calculation and the corresponding attention will be filled with zeros, such as row 3 and the last row in Fig.~\ref{fig_mask}.

\subsection{The KFDS mechanism}

\begin{figure}[h!]
	\centerline{\includegraphics[width=1.0\linewidth]{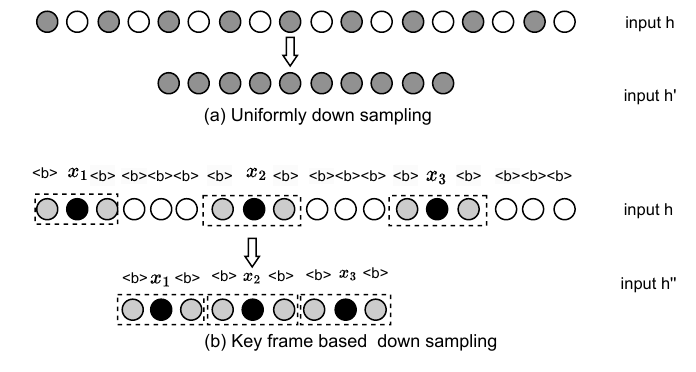}}
	\caption{(a) depicted the uniformly down sampling process. (b) depicted the KFDS mechanism. Here, all necessary frames are calculated based on the key frames and marked according to the index of the acoustic features, and then unnecessary frames are discarded.}\label{fig_kfds}
\end{figure}

The KFSA mechanism can only reduce the calculation of self-attention, it does not reduce the length of input sequences feeding into the second Conformer encoder. We further propose a key frame based downsampling mechanism, named KFDS. Different from KFSA, blank frames which do not locate at the neighbor of key frames will be discarded directly. Fig.~\ref{fig1}(c) gives a pipeline of KFDS-based models. Fig.~\ref{fig_kfds} depicts the downsampling procedure. After KFDS, the length of the input sequence reduces to $(2w+1)L$, where $L$ is the non-blank frame number of intermediate CTC. Generally, in a speech recognition task,  the length of the input sequence $T$ is much greater than the length of the output sequence, which equals $L$ approximately. Apart from the reduction of input length, as shown in Fig.~\ref{fig_kfds}, compared with the uniform downsampling, the KFDS mechanism can also keep crucial frames with the guidance of intermediate CTC.

\section{EXPERIMENTS}
\label{sec:guidelines}
\subsection{Corpus and  Experimental Configurations}

In this work, we verify our proposed KFSA-based Conformer encoder-decoder network on two open-source datasets: AISHELL-1 \cite{bu2017aishell} and LibriSpeech \cite{7178964}. In all of our experiments, 80-dimensional log Mel-filter bank (Fbank) features are extracted from a 25ms window with a 10ms frame shift. SpecAugment is used for acoustic feature augmentation \cite{park19e_interspeech}. To conduct modeling on the AiSHELL-1, a vocabulary consisting of 4233 labels that incorporate Chinese characters and other special characters is employed. For LibriSpeech, a vocabulary comprising 5002 word pieces and other special characters is utilized.

We adopt hybrid CTC and attention-based auto encoder-decoder (AED) architecture following Wenet recipe \cite{zhang2022wenet}. For the encoder, there are 12 Conformer blocks. For each block, the convolutional kernel size is 31, the number of attention heads is 4, and the hidden dimensions of the attention and FFN layer are 256 and 2048 respectively. The decoder consists of 6 Transformer blocks, where each block has 4 attention heads. As shown in Fig.~\ref{fig1}, all models are trained with intermediate CTC loss $L_{ctc1}$, final CTC loss $L_{ctc2}$ and decoder's cross-entropy loss $L_{ce}$. The final objective function of the proposed ASR framework is defined as:

\begin{equation}
\mathcal{L}=\beta_0 * (\alpha_0*\mathcal{L}_{ctc1}+ \alpha_1 * \mathcal{L}_{ctc2})+ \beta_1 * \mathcal{L}_{ce}.
\end{equation}
where $\alpha_0+\alpha_1=1$ and $\beta_0+\beta_1=1$. In all the experiments, we just set $\beta_0=0.3$ and $\beta_1=0.7$ empirically. As for training details, we follow the training recipes provided by Wenet. \footnote{The recipe on AISHELL-1 and LibriSpeech is publicly available in: \url{https://github.com/wenet-e2e/wenet/tree/main/examples/aishell/s0/conf}; \url{https://github.com/wenet-e2e/wenet/tree/main/examples/librispeech/s0/conf}} It will be easy for others to reproduce our experiments. Note that, in order to obtain a better initial intermediate CTC guidance, KFSA and KFDS are introduced after the first $N$ normal training epochs. $N$ is 40 for AISHELL-1 and Librispeech experiments.



\subsection{Overall Results}
\subsubsection{Results on AISHELL-1}

\begin{table}
\centering    

\caption{The overall result on AISHELL-1 with CER. Encoder1 shares the same structure for all experiments. E0-E4 show results of KFSA with different configurations and E5-E7 show results of KFDS.}
\setlength{\tabcolsep}{1mm}
\begin{tabular}{llllc}
\hline
Models                   & Condition                                                                        & Test  & Drop ratio \\ \hline
Conformer\cite{zhang2022wenet}       & All                                                                          & 4.75 & /           \\
B0                      & All                                                                          & \textbf{4.58} & /           \\
\textcolor{black}{B0 (InterCTC)}                      & All                                                                          & 6.70 & /           \\

Efficient Conformer\cite{burchi2021efficient}                      & All                                                                          & 4.64 & /           \\
\hline
\textcolor{black}{E0 (KFSA)}                            & {[}-3, +3{]}  + K    & 4.61 & /           \\
\textcolor{black}{E1 (KFSA)}                            & {[}-2, +2{]}  + K    & 4.62 & /           \\
\textcolor{black}{E2 (KFSA)}                            & {[}-1, +1{]}  + K    & 4.59 & /           \\
E3 (KFSA)                             &  K         & \textbf{4.58} & /           \\
E4 (KFSA)                            & {[}-3, +3{]}       & 4.86 & /           \\ \hline
E5 (KFDS)                              & {[}-3, +3{]}  + K   & 4.49 & 31.06\%     \\
E6 (KFDS)                             & {[}-2, +2{]}  + K  & \textbf{4.45} & 43.29\%     \\
E7 (KFDS)                             & {[}-1, +1{]} + K    & 4.52 & \textbf{64.78}\%    \\
\hline

\hline
\end{tabular}
\end{table}

\begin{table}[]
    \centering
    \caption{
investigate the appropriate layer to subsampling (after the intermediate CTC loss) }
    \label{tab:my_label}
    \begin{tabular}{ccc}
    \hline
Models & Intermediate CTC loss position  &Test   \\
    \hline

E7 (KFDS)                    & 6 & \textbf{4.52}      \\    
E8 (KFDS)                    & 4& 4.61       \\
E9 (KFDS)                    & 2  & 4.74       \\

\hline
    \end{tabular}
   
\end{table}

TABLE \uppercase\expandafter{\romannumeral1} shows the overall character error rate (CER) on the AISHELL-1 test set. During decoding, CTC prefix beam search is used to generate N-best candidates first and then rescored using a Transformer decoder. We only report the final results here after the Transformer decoder rescore. In Table I, the result of the vanilla Conformer model was from \cite{zhang2022wenet}. However, there is no intermediate CTC during model training in \cite{zhang2022wenet}. For a fair comparison, we add intermediate CTC loss at the 6th layer during model B0 training, which is our baseline model with 4.58\% CER. \textcolor{black}{B0 (InterCTC) shows CER of the intermediate CTC, which achieves a CER of 6.70\%. Although the intermediate CTC gives higher CER, most of the errors are substitutions, which will not affect the selection of key frame. } We also compare our methods with Efficient Conformer \cite{burchi2021efficient}, which downsampled
feature sequences uniformly.

Model E0-E4  are trained using our proposed KFSA mechanism with different configurations. [-3, +3] means w = 3 in Eq 3, and “+ K” means all other key frames are used during attention calculation. \textcolor{black}{E0-E3 have very similar results althouth they use different local information.} In order to explore the effect of local context information and global key frames, E3 is trained using key frames only while E4 is trained using local information only. The CER of E4 drops to 4.86\%, which also proves that key frames contain more useful information and are crucial for attention mechanisms.

Models E5-E7 are trained using the KFDS mechanism with different local temporal context widths of various $w$. Compared with the KFSA-based models E0-E4, the KFDS-based models obtain lower CER and less computational complexity. Especially, model E7 can get 4.52\% CER with more than 64\% blank frames discarded, which is better than Efficient Conformer too. Note that our KFDS mechanism did not entirely rely on key frames in experiments. This is because the model’s predicted sequence length when calculating CTC Loss, must be at least twice the length of the ground truth sequence, which will be explored in the future.

Note that during models E0-E7 training, KFSA or KFDS is applied after the 6th Conformer layer. In addition, we also investigate the impact of downsampling from different layers of the encoder on the results of our KFDS. It is obvious that the earlier we apply KFDS, the more computation can be reduced. In Table II, models E7-E9 are trained using the KFDS mechanism after the 6th, 4th, and 2nd Conformer layers respectively. From Table II, we can see that earlier downsampling will result in a little worse recognition performance, from 4.52\% (E7)to 4.74\% (E9). In practice, there is usually a tradeoff between CER and computational complexity. Even if we start downsampling after the second layer, we can still get an acceptable result, with only 4.8\% relative CER degradation compared with E7.

\begin{table}[]

\caption{The overall result on LibriSpeech with WER. For E6-E7, we have labeled the drop ratio of discarded frames after its WER}
\setlength{\tabcolsep}{1mm}
\begin{tabular}{llccc}
\hline
Models                   & Condition                                                                        & Test clean  & Test other \\ \hline
Conformer\cite{zhang2022wenet}     & All                                                                          & 3.18 & 8.72           \\
B0                      & All                                                                          & 3.14 & 8.26           \\ 
\textcolor{black}{B0 (InterCTC)}                      & All                                                                          & 4.34 & 12.03           \\ 
Efficient Conformer\cite{burchi2021efficient}                       & All                                                                          & 3.17 & 8.21           \\
\hline
E1 (KFSA)                           &  K  & 3.08 & 8.30   \\ \hline

\textcolor{black}{E6 (KFDS)}                           &  \begin{tabular}[c]{@{}l@{}}{[}-2, +2{]} \\+ K \end{tabular}  & 3.04 (53.15\%)& 7.85 (53.26\%)  \\

E7 (KFDS)                           &  \begin{tabular}[c]{@{}l@{}}{[}-1, +1{]} \\+ K \end{tabular}  & 3.09 (62.25\%)& 7.96 (62.61\%)   \\

\hline
\end{tabular}
\end{table}

\subsubsection{Results on LibriSpeech}
To verify the effectiveness of our proposed method, we also conducted experiments on the LibriSpeech dataset. In this section, we will report the results on the clean and other test sets, with the word error rate (WER) serving as the evaluation metric.

As shown in TABLE \uppercase\expandafter{\romannumeral3}, the baseline model B0 with intermediate CTC loss has better performance than the vanilla Conformer. \textcolor{black}{In addition, the intermediate CTC achieves 4.34\% and 12.03\% WER on the clean and other test sets.} The KFSA-based model achieves a WER of 3.08\% on Test clean and surpassing the vanilla Conformer 0.10\% absolutely, and overperforming the Efficient Conformer 0.09\%. On Test other sets, The KFSA-based model’s performance is comparable with B0 and Efficient Conformer. 
\textcolor{black}{As for the KFDS-based models, E6 achieves the lowest CER on both clean and other sets and E7 can drop more than 62\% frames with a little performance degradation. Note that, on Test other set, our best model E6 obtains 4.6\% relative CER reduction compared with Efficient Conformer, which proves that our proposed method can keep more crucial information and discard redundancy frames precisely.}

\section{CONCLUSION}

In this paper, inspired by previous CTC-guided blank skipping during decoding work, we introduce the key frame mechanism into the Conformer model. Specifically, we utilize intermediate CTC to generate key frames and drop blank frames, using key frames for the calculation in the subsequent Conformer blocks. Firstly, we propose KFSA, a novel method to reduce the computational complexity in the self-attention
mechanism. Experimental results on KFSA prove that key frames contain crucial information during attention calculation. The reduction of non-key frames will not affect the performance of ASR models. Then, we also introduce the KFDS mechanism to drop blank frames for the second encoder. Experimental results demonstrate the superiority of our approach over the baseline model and drop at least 60\% of frames. Future work will focus on the implementation of KFDS using only key frames to solve the problem of the limitation of the label input length and the speech sequence length of the CTC Loss function.

%
\vfill

\newpage
\balance
\bibliography{reference_google}

\begin{thebibliography}{10}
\providecommand{\url}[1]{#1}
\csname url@samestyle\endcsname
\providecommand{\newblock}{\relax}
\providecommand{\bibinfo}[2]{#2}
\providecommand{\BIBentrySTDinterwordspacing}{\spaceskip=0pt\relax}
\providecommand{\BIBentryALTinterwordstretchfactor}{4}
\providecommand{\BIBentryALTinterwordspacing}{\spaceskip=\fontdimen2\font plus
\BIBentryALTinterwordstretchfactor\fontdimen3\font minus
  \fontdimen4\font\relax}
\providecommand{\BIBforeignlanguage}[2]{{%
\expandafter\ifx\csname l@#1\endcsname\relax
\typeout{** WARNING: IEEEtran.bst: No hyphenation pattern has been}%
\typeout{** loaded for the language `#1'. Using the pattern for}%
\typeout{** the default language instead.}%
\else
\language=\csname l@#1\endcsname
\fi
#2}}
\providecommand{\BIBdecl}{\relax}
\BIBdecl

\bibitem{vaswani2017attention}
A.~Vaswani, N.~Shazeer, N.~Parmar, J.~Uszkoreit, L.~Jones, A.~N. Gomez,
  {\L}.~Kaiser, and I.~Polosukhin, ``Attention is all you need,''
  \emph{Advances in neural information processing systems}, vol.~30, 2017.

\bibitem{gulati2020conformer}
A.~Gulati, J.~Qin, C.-C. Chiu, N.~Parmar, Y.~Zhang, J.~Yu, W.~Han, S.~Wang,
  Z.~Zhang, Y.~Wu, and R.~Pang, ``{Conformer: Convolution-augmented Transformer
  for Speech Recognition},'' in \emph{Proc. Interspeech 2020}, 2020, pp.
  5036--5040.

\bibitem{2016Listen}
W.~Chan, N.~Jaitly, Q.~Le, and O.~Vinyals, ``Listen, attend and spell: A neural
  network for large vocabulary conversational speech recognition,'' in
  \emph{2016 IEEE international conference on acoustics, speech and signal
  processing (ICASSP)}.\hskip 1em plus 0.5em minus 0.4em\relax IEEE, 2016, pp.
  4960--4964.

\bibitem{2018Speech}
L.~Dong, S.~Xu, and B.~Xu, ``Speech-transformer: a no-recurrence
  sequence-to-sequence model for speech recognition,'' in \emph{2018 IEEE
  international conference on acoustics, speech and signal processing
  (ICASSP)}.\hskip 1em plus 0.5em minus 0.4em\relax IEEE, 2018, pp. 5884--5888.

\bibitem{graves2006connectionist}
A.~Graves, S.~Fern{\'a}ndez, F.~Gomez, and J.~Schmidhuber, ``Connectionist
  temporal classification: labelling unsegmented sequence data with recurrent
  neural networks,'' in \emph{Proceedings of the 23rd international conference
  on Machine learning}, 2006, pp. 369--376.

\bibitem{graves2012sequence}
A.~Graves, ``Sequence transduction with recurrent neural networks,''
  \emph{arXiv preprint arXiv:1211.3711}, 2012.

\bibitem{fan2021speech}
P.~FAN, X.~HUA, Y.~LIN, B.~YANG, J.~ZHANG, W.~GE, and D.~GUO, ``Speech
  recognition for air traffic control via feature learning and end-to-end
  training,'' \emph{IEICE Transactions on Information and Systems}, vol.
  E106.D, no.~4, pp. 538--544, 2023.

\bibitem{zhang2021tiny}
Y.~Zhang, S.~Sun, and L.~Ma, ``Tiny transducer: A highly-efficient speech
  recognition model on edge devices,'' in \emph{ICASSP 2021-2021 IEEE
  International Conference on Acoustics, Speech and Signal Processing
  (ICASSP)}.\hskip 1em plus 0.5em minus 0.4em\relax IEEE, 2021, pp. 6024--6028.

\bibitem{graves2014towards}
A.~Graves and N.~Jaitly, ``Towards end-to-end speech recognition with recurrent
  neural networks,'' in \emph{International conference on machine
  learning}.\hskip 1em plus 0.5em minus 0.4em\relax PMLR, 2014, pp. 1764--1772.

\bibitem{he2019streaming}
Y.~He, T.~N. Sainath, R.~Prabhavalkar, I.~McGraw, R.~Alvarez, D.~Zhao,
  D.~Rybach, A.~Kannan, Y.~Wu, R.~Pang \emph{et~al.}, ``Streaming end-to-end
  speech recognition for mobile devices,'' in \emph{ICASSP 2019-2019 IEEE
  International Conference on Acoustics, Speech and Signal Processing
  (ICASSP)}.\hskip 1em plus 0.5em minus 0.4em\relax IEEE, 2019, pp. 6381--6385.

\bibitem{battenberg2017exploring}
E.~Battenberg, J.~Chen, R.~Child, A.~Coates, Y.~G.~Y. Li, H.~Liu, S.~Satheesh,
  A.~Sriram, and Z.~Zhu, ``Exploring neural transducers for end-to-end speech
  recognition,'' in \emph{2017 IEEE automatic speech recognition and
  understanding workshop (ASRU)}.\hskip 1em plus 0.5em minus 0.4em\relax IEEE,
  2017, pp. 206--213.

\bibitem{chen2016phone}
Z.~Chen, W.~Deng, T.~Xu, and K.~Yu, ``Phone synchronous decoding with ctc
  lattice.'' in \emph{Interspeech}, 2016, pp. 1923--1927.

\bibitem{tian2021fsr}
Z.~Tian, J.~Yi, Y.~Bai, J.~Tao, S.~Zhang, and Z.~Wen, ``Fsr: Accelerating the
  inference process of transducer-based models by applying fast-skip
  regularization,'' \emph{arXiv preprint arXiv:2104.02882}, 2021.

\bibitem{xu2022multiblank}
H.~Xu, F.~Jia, S.~Majumdar, S.~Watanabe, and B.~Ginsburg, ``Multi-blank
  transducers for speech recognition,'' 2022.

\bibitem{yang2023blank}
Y.~Yang, X.~Yang, L.~Guo, Z.~Yao, W.~Kang, F.~Kuang, L.~Lin, X.~Chen, and
  D.~Povey, ``Blank-regularized ctc for frame skipping in neural transducer,''
  \emph{arXiv preprint arXiv:2305.11558}, 2023.

\bibitem{macoskey2021amortized}
J.~Macoskey, G.~P. Strimel, J.~Su, and A.~Rastrow, ``Amortized neural networks
  for low-latency speech recognition,'' \emph{arXiv preprint arXiv:2108.01553},
  2021.

\bibitem{wang2023accelerating}
Y.~Wang, Z.~Chen, C.~Zheng, Y.~Zhang, W.~Han, and P.~Haghani, ``Accelerating
  rnn-t training and inference using ctc guidance,'' in \emph{ICASSP 2023-2023
  IEEE International Conference on Acoustics, Speech and Signal Processing
  (ICASSP)}.\hskip 1em plus 0.5em minus 0.4em\relax IEEE, 2023, pp. 1--5.

\bibitem{10096467}
Y.~Yang, Y.~Pan, J.~Yin, J.~Han, L.~Ma, and H.~Lu, ``Hybridformer: Improving
  squeezeformer with hybrid attention and nsr mechanism,'' in \emph{ICASSP 2023
  - 2023 IEEE International Conference on Acoustics, Speech and Signal
  Processing (ICASSP)}, 2023, pp. 1--5.

\bibitem{burchi2021efficient}
M.~Burchi and V.~Vielzeuf, ``Efficient conformer: Progressive downsampling and
  grouped attention for automatic speech recognition,'' in \emph{2021 IEEE
  Automatic Speech Recognition and Understanding Workshop (ASRU)}.\hskip 1em
  plus 0.5em minus 0.4em\relax IEEE, 2021, pp. 8--15.

\bibitem{kim2022squeezeformer}
S.~Kim, A.~Gholami, A.~Shaw, N.~Lee, K.~Mangalam, J.~Malik, M.~W. Mahoney, and
  K.~Keutzer, ``Squeezeformer: An efficient transformer for automatic speech
  recognition,'' \emph{arXiv preprint arXiv:2206.00888}, 2022.

\bibitem{interctc}
J.~Lee and S.~Watanabe, ``Intermediate loss regularization for ctc-based speech
  recognition,'' in \emph{ICASSP 2021 - 2021 IEEE International Conference on
  Acoustics, Speech and Signal Processing (ICASSP)}, 2021, pp. 6224--6228.

\bibitem{bu2017aishell}
H.~Bu, J.~Du, X.~Na, B.~Wu, and H.~Zheng, ``Aishell-1: An open-source mandarin
  speech corpus and a speech recognition baseline,'' in \emph{2017 20th
  conference of the oriental chapter of the international coordinating
  committee on speech databases and speech I/O systems and assessment
  (O-COCOSDA)}.\hskip 1em plus 0.5em minus 0.4em\relax IEEE, 2017, pp. 1--5.

\bibitem{7178964}
V.~Panayotov, G.~Chen, D.~Povey, and S.~Khudanpur, ``Librispeech: An asr corpus
  based on public domain audio books,'' in \emph{2015 IEEE International
  Conference on Acoustics, Speech and Signal Processing (ICASSP)}, 2015, pp.
  5206--5210.

\bibitem{park19e_interspeech}
D.~S. Park, W.~Chan, Y.~Zhang, C.-C. Chiu, B.~Zoph, E.~D. Cubuk, and Q.~V. Le,
  ``{SpecAugment: A Simple Data Augmentation Method for Automatic Speech
  Recognition},'' in \emph{Proc. Interspeech 2019}, 2019, pp. 2613--2617.

\bibitem{zhang2022wenet}
B.~Zhang, D.~Wu, Z.~Peng, X.~Song, Z.~Yao, H.~Lv, L.~Xie, C.~Yang, F.~Pan, and
  J.~Niu, ``Wenet 2.0: More productive end-to-end speech recognition toolkit,''
  \emph{arXiv preprint arXiv:2203.15455}, 2022.

\end{thebibliography}

\end{document}